\documentclass[english,prl,reprint]{revtex4-1}
\usepackage[T1]{fontenc}
\usepackage[latin9]{inputenc}
\usepackage{graphicx}
\usepackage{esint}
\usepackage{babel}
\begin{document}

\title{Entropy-reducing dynamics of a double demon}

\author{Ian J. Ford\textsuperscript{1} and Michael Maitland\textsuperscript{2}}

\affiliation{\textsuperscript{1}Department of Physics and Astronomy and London
Centre for Nanotechnology, UCL, Gower Street, London WC1E 6BT, U.K.}

\affiliation{\textsuperscript{2}Centre for Complexity Science, University of
Warwick, Coventry CV4 7AL, U.K.}
\begin{abstract}
We study the reduction in total entropy, and associated conversion
of environmental heat into work, arising from the coupling and decoupling
of two systems followed by processing determined by suitable mutual
feedback. The scheme is based on the actions of Maxwell's demon, namely
the performance of a measurement on a system followed by an exploitation
of the outcome to extract work. When this is carried out in a symmetric
fashion, with each system informing the exploitation of the other
(and both therefore acting as a demon), it may be shown that the second
law can be broken, a consequence of the self-sorting character of
the system dynamics.
\end{abstract}
\maketitle

Entropy production may be viewed as the irreversible growth in uncertainty
of the microscopic state of a system, arising from the complexity
of the underlying dynamics and the incompleteness of the model we
employ to represent it. It can be quantified using a framework of
stochastic rules describing the forward evolution in time of a system
influenced by its coarse grained environment \cite{seifertprinciples}.
Nevertheless, it is possible to conceive of procedures, often involving
a time-asymmetric feedback mechanism, that operate against the usual
tendency for processes to operate irreversibly, a classic example
being Maxwell's demon \cite{Leff03}. The argument is that a measurement
of a system can reveal a route for exploitation that leads to a reduction
in entropy; typically the conversion of environmental heat to potential
energy represented by the raising of a weight.

The original demon exploited his observations to sort molecules of
a gas into fast and slow groups, creating a resource for a heat engine
without the expenditure of work. Szilard's later conception of a demon-operated
heat engine made this explicit \cite{Szilard29}. Maxwell saw no edict
against breaking the second law by time-asymmetric dynamical processing
\cite{Maxwell71,Earman98,Earman99,HemShen12}, but much attention
has been given to finding a way to `exorcise the demon' and protect
the second law in these circumstances. The majority view is that dissipative
processes, operating either in the act of measurement \cite{Szilard29,Brillouin51,Sagawa09,Granger11,Mandal12,Sagawa12,Sagawa12b,Ford16}
or the act of restoring the initial condition of the demon prior to
measurement \cite{Landauer61,Bennett73,Bennett82,Plenio01}, generate
enough entropy to cancel out any possible gains \cite{Abreu11,Barato13}.

But we can imagine dynamical schemes that emulate a successful demon
and we are obliged either to accept that they are possible, or to
find reasons to exclude them. Consider, for example, a particle tethered
by a harmonic spring to a point and coupled to a heat bath. The tether
point might be moved instantaneously towards the current position
of the particle, relaxing the spring and harvesting potential energy
to lift a weight. Subsequently, the system will evolve back towards
equilibrium under the influence of the heat bath, with its mean potential
energy replenished through the absorption of heat. There is no act
of measurement and no demon here, just an autonomous dynamical system
that employs feedback, breaking time reversal symmetry. We might call
such dynamics self-sorting and this system a self-adjusting oscillator.

On the other hand, we might feel uneasy about a system that alters
the dynamical rules that control its future behaviour, depending on
its current state, and question whether such a system is admissible
for thermodynamic consideration. We might prefer to channel the feedback
instead through a second entity, a demon, such that the system dynamics
cannot be described as self-sorting. The system would be coupled to
the demon, or more prosaically a measuring device, in such a way that
establishes a correlation between their coordinates, and then decoupled.
The subsequent exploitation of the system would then be determined
by the device coordinate. Since the effects of the feedback are felt
by the system \emph{after} the device is decoupled, there is no element
of self-sorting in the dynamics.

Analysis of such a procedure indeed shows that the second law is preserved.
The process of coupling and decoupling must be entropy generating
if it is to yield an exploitable correlation between system and device
\cite{Granger11,Sagawa12,Ford16}. This entropy production exceeds
the reduction in entropy made possible by the measurement, and so
overall the entropy is never decreased. This can be regarded as a
satisfactory outcome.

However, if we accept that the state of a device can inform the subsequent
exploitation of the system, it is natural to consider a symmetric
arrangement where the state of the system is allowed to inform an
exploitation of the device. Both parts act as a demon and we might
therefore call this a realisation of a `double demon'. Intuitively,
we suspect that the combination might display self-sorting behaviour.
The feedback mechanism preserves the second law when applied in a
demon-system context, but when operated mutually the effect might
be different.

Our purpose is to analyse the dynamics and thermodynamics of a double
demon constructed using harmonic springs, and to investigate its irreversibility.
We find that a suitable exploitation protocol makes possible an overall
entropy reduction.

We consider two 1-d harmonic oscillators that can be coupled and decoupled
through a further harmonic spring. Both oscillators are influenced
by noise from the environment. We consider four intervals of time.
In the period $-\infty\le t\le-\tau_{{\rm m}}$ the coupling spring
strength $K$ is zero, the oscillator spring strengths are unity.
We take $kT$ to equal unity so at $t=-\tau_{{\rm m}}$ an equilibrium
state is established described by the probability density function
(pdf) $p(x,y,-\tau_{{\rm m}})=(2\pi)^{-1}\exp[-(x^{2}+y^{2})/2]$
where $x$ and $y$ are the oscillator displacements.

\begin{figure}
\begin{centering}
\includegraphics[width=1\columnwidth]{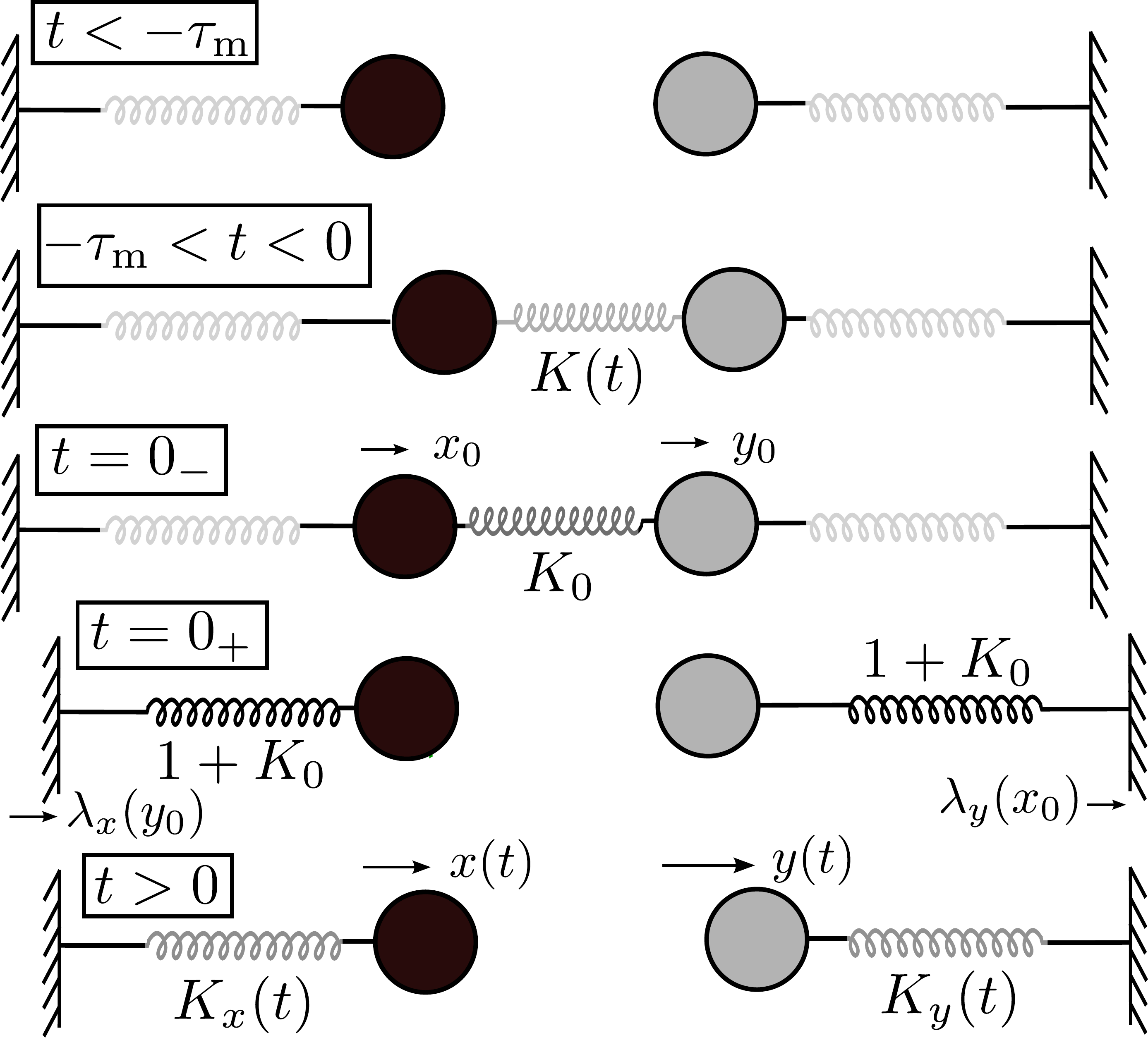}
\par\end{centering}

\centering{}\caption{Two oscillators in a noisy environment are coupled and decoupled in
the period $-\tau_{{\rm m}}\le t\le0$ using a harmonic spring of
strength $K(t)$. The spring strengths $K_{x,y}$ are adjusted and
tether points $\lambda_{x,y}$ shifted, using feedback from the displacements
$x_{0}$ and $y_{0}$ at $t=0$, in preparation for an exploitation
procedure in the period $t>0$. \label{fig:system+device}}
\end{figure}

In the next time interval $-\tau_{{\rm m}}\le t\le0$ the oscillators
interact through a nonzero $K(t)$. The situation is illustrated in
Fig. \ref{fig:system+device}, where the shading of the springs indicates
their strength. We consider the coupling spring constant to evolve
to a value $K(0)=K_{0}$ just prior to $t=0$ (labelled $t=0_{-}$)
and then to go to zero abruptly. We model $x$ and $y$ in this period
using overdamped stochastic differential equations
\begin{eqnarray}
dx & = & -xdt-K(t)(x-y)dt+\sqrt{2}\,dW_{x},\nonumber \\
dy & = & -ydt-K(t)(y-x)dt+\sqrt{2}\,dW_{y},\label{eq:1}
\end{eqnarray}
representing the effects of the various spring forces, with white
environmental noise described by increments $dW_{x}$ and $dW_{y}$
in separate Wiener processes. The work performed during this `measurement'
period is
\begin{eqnarray}
W_{{\rm m}} & = & \int_{-\tau_{{\rm m}}}^{0}\frac{1}{2}\frac{dK}{dt}[x(t)-y(t)]^{2}dt.\label{eq:6}
\end{eqnarray}

The pdf satisfying the Fokker-Planck equation in the measurement interval
takes the form
\begin{equation}
\!p(x,y,t)\!=\!\frac{[1+2\tilde{K}]^{\frac{1}{2}}}{2\pi}\exp\!\left[-\frac{x^{2}}{2}-\frac{y^{2}}{2}-\frac{\tilde{K}(y-x)^{2}}{2}\right]\!,\!\!\label{eq:2}
\end{equation}
with the time-dependent parameter $\tilde{K}(t)$ determined by
\begin{equation}
d\tilde{K}/dt=-2(\tilde{K}-K)(1+2\tilde{K}),\label{eq:33b}
\end{equation}
with $\tilde{K}(-\tau_{{\rm m}})=0$. In the quasistatic limit, $\tilde{K}(t)$
follows the evolution of the coupling strength $K(t)$ exactly. The
stochastic entropy production during measurement is given by \cite{seifertprinciples,SpinneyFord12b,Ford16,fordbook}
\begin{eqnarray}
d\Delta s_{{\rm tot}}^{{\rm m}} & = & (\tilde{K}-K)\Bigl[4dt-(1+2\tilde{K})(y-x)^{2}dt\nonumber \\
 &  & +(x-y)(dx-dy)\Bigr],\label{eq:4}
\end{eqnarray}
which implies a mean rate of production
\begin{equation}
d\langle\Delta s_{{\rm tot}}^{{\rm m}}\rangle/dt=4(\tilde{K}-K)^{2}/(1+2\tilde{K}).\label{eq:5}
\end{equation}

At $t=0$, the joint pdf may be represented as $p(x_{0},y_{0},0)=p(x_{0},0\vert y_{0})p(y_{0},0)$
where $x_{0}=x(0)$ and $y_{0}=y(0)$, with the state of the first
oscillator described by a conditional pdf
\begin{eqnarray}
p(x_{0},0\vert y_{0}) & = & [(1+\tilde{K}_{0})/2\pi]^{1/2}\nonumber \\
 & \times & \exp\left[\!-\frac{(1+\tilde{K}_{0})}{2}\left(x_{0}-\frac{\tilde{K}_{0}y_{0}}{1+\tilde{K}_{0}}\right)^{2}\right],\quad\label{eq:7b-1}
\end{eqnarray}
where $\tilde{K}_{0}=\tilde{K}(0)$, and the second oscillator described
by
\begin{equation}
p(y_{0},0)=\left(\kappa/2\pi\right)^{1/2}\exp(-\kappa y_{0}^{2}/2),\label{eq:7b}
\end{equation}
with $\kappa=(1+2\tilde{K}_{0})/(1+\tilde{K}_{0})$.

If the measurement process took the form of quasistatic coupling and
instantaneous decoupling (denoted `qi'), for which $\tilde{K}_{0}$
would equal $K_{0}$, the mean work of measurement would be a free
energy of coupling minus the mean energy of the coupling spring at
$t=0$, which is
\begin{equation}
\langle W_{{\rm m}}^{{\rm qi}}\rangle=\frac{1}{2}\ln(1+2K_{0})-\frac{K_{0}}{1+2K_{0}}.\label{eq:7}
\end{equation}

In the third period $0\le t\le\tau_{{\rm e}}$ the outcome of the
measurement is exploited through changes in the strengths $K_{x}(t)$
and $K_{y}(t)$ of the oscillator springs and the positions $\lambda_{x}$
and $\lambda_{y}$ of their tethering points. Optimal exploitation
sequences for a single oscillator after measurement have been studied
previously \cite{Abreu11,Granger11,Sagawa12}. We similarly employ
$\lambda_{x}=K_{0}y_{0}/(1+K_{0})$, $\lambda_{y}=K_{0}x_{0}/(1+K_{0})$,
and $K_{x}(0)=K_{y}(0)=1+K_{0}$, the rationale for which will become
clear, and $K_{x}(\tau_{{\rm e}})=K_{y}(\tau_{{\rm e}})=1$. The changes
at the start of the `exploitation' interval are indicated in Fig.
\ref{fig:system+device} (labelled $t=0_{+}$). The subsequent dynamics
are modelled using
\begin{eqnarray}
dx & = & -K_{x}(t)(x-\lambda_{x})dt+\sqrt{2}\,dW_{x},\nonumber \\
dy & = & -K_{y}(t)(y-\lambda_{y})dt+\sqrt{2}\,dW_{y},\label{eq:8}
\end{eqnarray}
and the work of exploitation is given by
\begin{equation}
W_{{\rm e}}=\int_{0}^{\tau_{{\rm m}}}\frac{1}{2}\left[\frac{dK_{x}}{dt}\left(x(t)-\lambda_{x}\right)^{2}+\frac{dK_{y}}{dt}\left(y(t)-\lambda_{y}\right)^{2}\right]dt.\label{eq:8a}
\end{equation}
If the pdf $p(x_{0},y_{0},0)$ is characterised by $\tilde{K}_{0}=K_{0}$
and if instantaneous changes to the spring strengths and tether points
are followed by a quasistatic evolution of $K_{x}$ and $K_{y}$ (a
process denoted `iq'), then the mean work of exploitation is given
by
\begin{equation}
\langle W_{{\rm e}}^{{\rm iq}}\rangle=\frac{K_{0}}{2(1+2K_{0})}-\frac{1}{2}\ln(1+K_{0}),\label{eq:9}
\end{equation}
for each oscillator. The mean work of measurement and exploitation
for a protocol consisting of quasistatic coupling, instantaneous decoupling
and changes in spring parameters, followed by further quasistatic
processing (denoted `qiiq') is therefore
\begin{equation}
\!\langle W_{{\rm m+e}}^{{\rm qiiq},1}\rangle=\langle W_{{\rm m}}^{{\rm qi}}\rangle+\langle W_{{\rm e}}^{{\rm iq}}\rangle=\!\frac{1}{2}\ln\!\left[\frac{1+2K_{0}}{1+K_{0}}\right]-\frac{K_{0}}{2(1+2K_{0})},\label{eq:10}
\end{equation}
for exploitation of just the first oscillator (i.e. where $K_{y}$
and $\lambda_{y}$ are not modified for the exploitation interval).
Since this mean work cannot be negative, the second law is preserved.
However, if both oscillators are subjected to the exploitation procedure
we get
\begin{equation}
\langle W_{{\rm m+e}}^{{\rm qiiq},2}\rangle=\langle W_{{\rm m}}^{{\rm qi}}\rangle+2\langle W_{{\rm e}}^{{\rm iq}}\rangle=\frac{1}{2}\ln\left[\frac{1+2K_{0}}{(1+K_{0})^{2}}\right]\le0,\label{eq:11}
\end{equation}
which offers the prospect of a violation of the law.

To check such a claim, we compute the stochastic entropy production
during exploitation. We write the pdf as $p(x,y,t)=p(x,t\vert y_{0})p(y,t\vert x_{0})$,
such that at $t=0$ the conditional pdf $p(x,t\vert y_{0})$ takes
the form in Eq. (\ref{eq:7b-1}) and $p(y,t\vert x_{0})$ is given
by Eq. (\ref{eq:7b}). This asymmetric specification of the pdf provides
a rationale for the chosen exploitation protocol. If a quasistatic-instantaneous
(qi) measurement procedure has been performed, such that $\tilde{K}_{0}=K_{0}$,
then the changes made to $K_{x}$ and $\lambda_{x}$ at $t=0$ effectively
place the first oscillator in a state of canonical equilibrium. A
subsequent quasistatic evolution of the spring strength of the first
oscillator could be carried out without the generation of entropy.

In general, however, the evolving pdf of the first oscillator would
be written
\begin{equation}
p(x,t\vert y_{0})\!=\![\tilde{K}_{x}(t)/2\pi]^{1/2}\exp(-\tilde{K}_{x}(t)[x-\bar{x}(t)]^{2}/2),\,\!\!\!\label{eq:13}
\end{equation}
with time-dependent parameters $\tilde{K}_{x}(t)$ and $\bar{x}(t)$
given by
\begin{eqnarray}
d\tilde{K}_{x}/dt & = & -2\tilde{K}_{x}(\tilde{K}_{x}-K_{x}),\label{eq:14}\\
d\bar{x}/dt & = & -K_{x}[\bar{x}-\lambda_{x}(y_{0})],\label{eq:14a}
\end{eqnarray}
subject to $\tilde{K}_{x}(0)=1+\tilde{K}_{0}$ and $\bar{x}(0)=\tilde{K}_{0}y_{0}/(1+\tilde{K}_{0})$.
The stochastic entropy production associated with the first oscillator
during exploitation is \cite{SpinneyFord12b}
\begin{eqnarray}
 &  & d\Delta s_{{\rm tot}}^{x}=(2-\tilde{K}_{x}[x-\bar{x}]^{2})(\tilde{K}_{x}-K_{x})dt+K_{x}\lambda_{x}dx\nonumber \\
 &  & -\tilde{K}_{x}\bar{x}dx\!+\!\tilde{K}_{x}(x-\bar{x})K_{x}(\bar{x}-\lambda_{x})dt\!-\!(K_{x}-\tilde{K}_{x})xdx.\qquad\label{eq:20-1}
\end{eqnarray}
For a qi measurement process $\tilde{K}_{0}=K_{0}$ and $\bar{x}(0)=\lambda_{x}$,
in which case $d\Delta s_{{\rm tot}}^{x}$ reduces to
\begin{equation}
\!d\Delta s_{{\rm tot}}^{{\rm qi},x}\!\!=\!(\tilde{K}_{x}-K_{x})(2dt-\tilde{K}_{x}[x-\lambda_{x}]^{2}dt+[x-\lambda_{x}]dx),\label{eq:15}
\end{equation}
leading to an average rate of production
\begin{equation}
d\langle\Delta s_{{\rm tot}}^{{\rm qi},x}\rangle/dt=(\tilde{K}_{x}-K_{x})^{2}/\tilde{K}_{x}.\label{eq:16}
\end{equation}
Similarly, the pdf of the second oscillator may be written
\begin{equation}
\!p(y,t\vert x_{0})=[\tilde{K}_{y}(t)/2\pi]^{1/2}\exp(-\tilde{K}_{y}(t)[y-\bar{y}(t)]^{2}/2),\!\!\!\!\label{eq:17}
\end{equation}
for the exploitation interval, with
\begin{eqnarray}
d\tilde{K}_{y}/dt & = & -2\tilde{K}_{y}(\tilde{K}_{y}-K_{y}),\label{eq:18}\\
d\bar{y}/dt & = & -K_{y}[\bar{y}-\lambda_{y}(x_{0})],\label{eq:19}
\end{eqnarray}
but this time subject to $\tilde{K}_{y}(0)=(1+2\tilde{K}_{0})/(1+\tilde{K}_{0})$
and $\bar{y}(0)=0$. The stochastic entropy production associated
with the second oscillator during exploitation is \cite{SpinneyFord12b}

\begin{eqnarray}
 &  & d\Delta s_{{\rm tot}}^{y}\!=\!(2-\tilde{K}_{y}[y-\bar{y}]^{2})(\tilde{K}_{y}-K_{y})dt\!+\!K_{y}\lambda_{y}dy\nonumber \\
 &  & -\tilde{K}_{y}\bar{y}dy\!+\!\tilde{K}_{y}(y-\bar{y})K_{y}(\bar{y}-\lambda_{y})dt\!-\!(K_{y}-\tilde{K}_{y})ydy.\qquad\label{eq:20}
\end{eqnarray}

It is instructive to consider a special case where we assume $\tilde{K}_{0}=K_{0}$,
set $K_{y}$ equal to $1+K_{0}$ for the initial part of the exploitation
interval, and compute a stochastic entropy production $d\Delta s_{{\rm tot}}^{{\rm rel},y}$
associated with the relaxation of the parameters $\tilde{K}_{y}$
and $\bar{y}$ to $K_{y}$ and $\lambda_{y}(x_{0})$, respectively.
This is the key to understanding the breakage of the second law. The
mean rate of production, averaged over $x_{0}$ and $y_{0}$, is given
by

\begin{equation}
d\langle\Delta s_{{\rm tot}}^{{\rm rel},y}\rangle/dt=-K_{0}^{2}\tilde{K}_{y}^{{\rm rel}}\exp[-2(1+K_{0})t]/(1+2K_{0}),\label{eq:21}
\end{equation}
with $d\tilde{K}_{y}^{{\rm rel}}/dt=-2\tilde{K}_{y}^{{\rm rel}}(\tilde{K}_{y}^{{\rm rel}}-1-K_{0})$.
Note that this mean rate of production is \emph{negative. }After this
relaxation, $d\langle\Delta s_{{\rm tot}}^{y}\rangle/dt$ takes the
form of an analogue of Eq. (\ref{eq:16}) associated with the deviation
between $\tilde{K}_{y}$ and $K_{y}$ over the remainder the exploitation
interval.

In the final period $\tau_{{\rm e}}\le t\le\infty$ the oscillator
spring constants have returned to unity and the system relaxes until
there is no further stochastic entropy production. The shifts in the
tether points are irrelevant to the irreversibility, since a quasistatic
process can take the $\lambda_{x,y}$ back to zero at no cost in mean
work or mean entropy production. This completes the cycle.

The only contribution to the mean total entropy production for a qiiq
process involving exploitation of both oscillators is the relaxation
represented by Eq. (\ref{eq:21}). When integrated over time this
gives
\begin{equation}
\langle\Delta s_{{\rm tot}}^{{\rm rel},y}\rangle=\langle\Delta s_{{\rm tot}}^{{\rm qiiq},2}\rangle=\frac{1}{2}\ln\left[\frac{1+2K_{0}}{(1+K_{0})^{2}}\right]\le0,\label{eq:22}
\end{equation}
which is consistent with the negative mean work $\langle W_{{\rm m+e}}^{{\rm qiiq},2}\rangle$
in Eq. (\ref{eq:11}) obtained from a cycle of measurement and exploitation
under such conditions.

If exploitation is \emph{not} invoked for the second oscillator, its
relaxation would take place while $K_{y}$ remained equal to unity
and with $\lambda_{y}=0$. The mean total entropy production for this
case can be shown to be consistent with the positive mean work $\langle W_{{\rm m+e}}^{{\rm qiiq},1}\rangle$
in Eq. (\ref{eq:10}). Thus the mechanical and entropic assessments
of irreversibility are consistent with one another for the special
case of the qiiq process.

If the double demon oscillator system is processed slowly enough,
we therefore expect environmental heat to be converted into stored
potential energy, when averaged over many realisations, with a maximum
change in the total thermodynamic entropy per cycle given by Eq. (\ref{eq:22}).
This arises from the exploitation of each oscillator in a manner informed
by the state of the other at $t=0$, a symmetrising of the feedback
traditionally employed in a system-demon scenario. We next study how
this might emerge in numerical simulations.

\begin{figure}
\begin{centering}
\includegraphics[width=1\columnwidth]{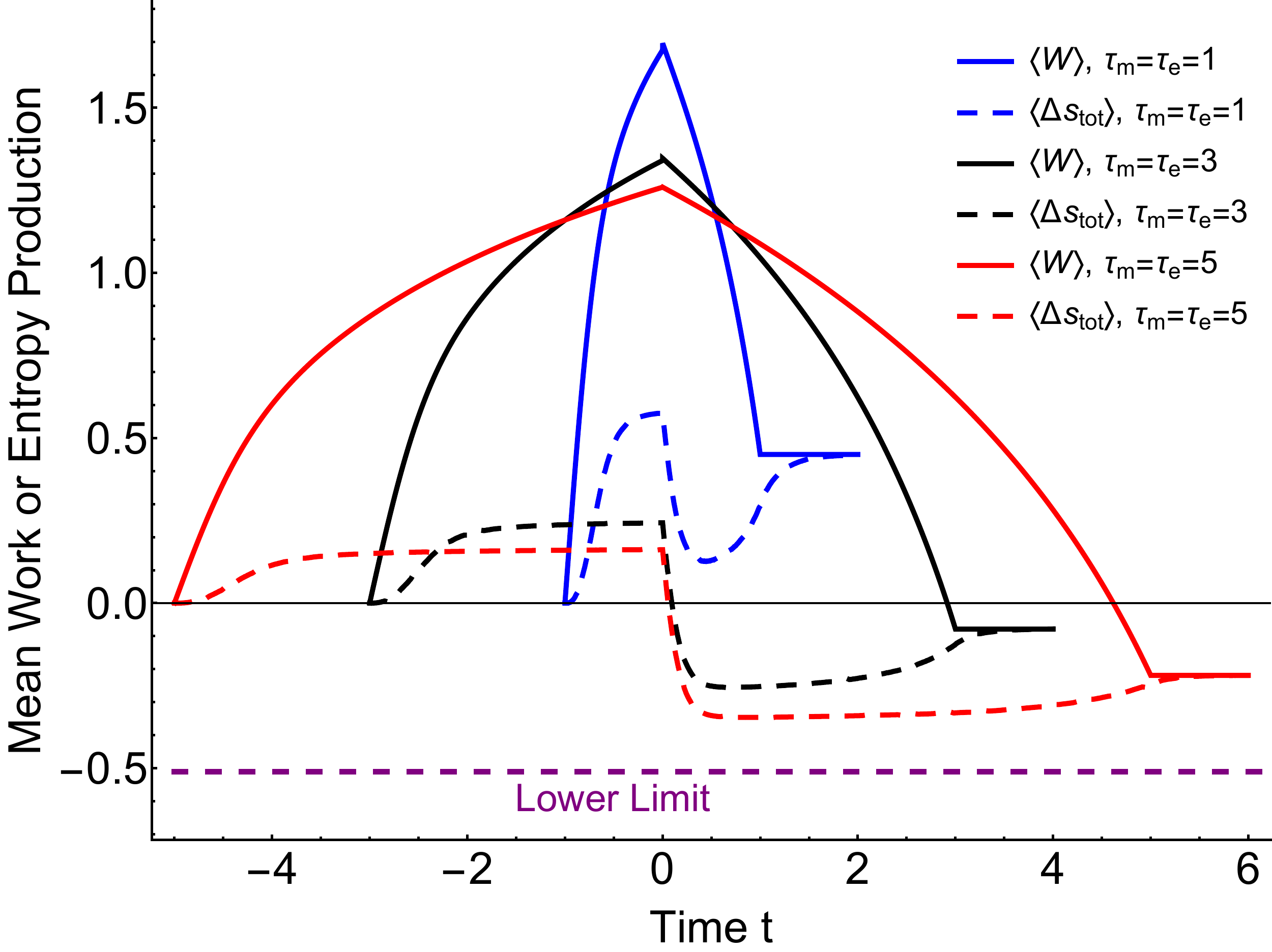}
\par\end{centering}

\caption{Evolution of the mean work performed on the two oscillator system
(solid lines), and the associated mean stochastic entropy production
(dashed lines), obtained from numerical simulations of the process
specified in the main text, conducted over a range of timescales $\tau=\tau_{{\rm \ensuremath{m}}}=\tau_{{\rm e}}$.
The quantities become equal when equilibrium is restored at the end
of the process, and they approach a negative limit when the process
timescale goes to infinity. \label{fig:numerics.}}
\end{figure}

Equations (\ref{eq:1}) and (\ref{eq:8}) for the stochastic dynamics
of the oscillators, (\ref{eq:6}) and (\ref{eq:8a}) for the performance
of work and (\ref{eq:4}), (\ref{eq:20-1}) and (\ref{eq:20}) for
the production of stochastic entropy have been solved numerically
for the following protocols of evolution of the spring constants:
$K(t)=K_{0}(1+t/\tau_{{\rm m}})$ for $-\tau_{{\rm m}}\le t\le0$,
otherwise zero; $K_{x,y}(t)=(1+K_{0})-K_{0}t/\tau_{{\rm e}}$ for
$0\le t\le\tau_{{\rm e}}$, otherwise unity; and tether positions
$\lambda_{x}=K_{0}y_{0}/(1+K_{0})$ and $\lambda_{y}=K_{0}x_{0}/(1+K_{0})$
for $t\ge0$, otherwise zero. Equations (\ref{eq:33b}), (\ref{eq:14}),
(\ref{eq:14a}), (\ref{eq:18}) and (\ref{eq:19}) are solved in the
appropriate time intervals to provide the necessary parameters $\tilde{K}(t)$,
$\tilde{K}_{x}(t)$, $\bar{x}(t)$, $\tilde{K}_{y}(t)$, and $\bar{y}(t)$.

We choose $K_{0}=4$ and generate sets of realisations for various
values of $\tau_{{\rm m}}$ and $\tau_{{\rm e}}$. The averages of
work done and stochastic entropy produced are shown in Fig. \ref{fig:numerics.}
using a timestep of $5\times10^{-4}$ and $10^{6}$ trajectories for
each case. The two quantities coincide at the end of the cycle, as
the laws of thermodynamics suggest they should, and for slower processes
they approach the negative limit given by Eqs. (\ref{eq:11}) and
(\ref{eq:22}). The reduction in entropy is achieved through the feedback
invoked at $t=0$ and the partial harvesting of the potential energy
of the springs. However, positive mean entropy production arises from
the deviations of $\tilde{K}$, $\tilde{K}_{x}$ and $\tilde{K}_{y}$
from $K$, $K_{x}$ and $K_{y}$, respectively, and this occurs more
strongly for faster processes.

We have shown that the stochastic evolution of two oscillators evolving
according to a particular form of autonomous time-asymmetric dynamics
can break the second law. Their interactions are conceived as an extension
of a scenario where a measurement made by a device or demon is used
to inform the exploitation of a system in order to convert environmental
heat into work. In that case the second law is preserved because of
a prior work of measurement, but passing feedback in both directions,
thus allowing the device to be manipulated in a fashion informed by
the system, extracts additional work. The total thermodynamic entropy
can be reduced after completion of such a sequence of measurement
and exploitation, and we attribute this to the self-sorting dynamics
of the double demon.

We thank Stefan Grosskinsky and Rosemary J. Harris for helpful discussions,
and acknowledge support from Engineering and Physical Sciences Research
Council (EPSRC) Grant No. EP/I01358X/1, and the COST1209 network.


\begin{thebibliography}{22}%
\makeatletter
\providecommand \@ifxundefined [1]{%
 \@ifx{#1\undefined}
}%
\providecommand \@ifnum [1]{%
 \ifnum #1\expandafter \@firstoftwo
 \else \expandafter \@secondoftwo
 \fi
}%
\providecommand \@ifx [1]{%
 \ifx #1\expandafter \@firstoftwo
 \else \expandafter \@secondoftwo
 \fi
}%
\providecommand \natexlab [1]{#1}%
\providecommand \enquote  [1]{``#1''}%
\providecommand \bibnamefont  [1]{#1}%
\providecommand \bibfnamefont [1]{#1}%
\providecommand \citenamefont [1]{#1}%
\providecommand \href@noop [0]{\@secondoftwo}%
\providecommand \href [0]{\begingroup \@sanitize@url \@href}%
\providecommand \@href[1]{\@@startlink{#1}\@@href}%
\providecommand \@@href[1]{\endgroup#1\@@endlink}%
\providecommand \@sanitize@url [0]{\catcode `\\12\catcode `\$12\catcode
  `\&12\catcode `\#12\catcode `\^12\catcode `\_12\catcode `\%12\relax}%
\providecommand \@@startlink[1]{}%
\providecommand \@@endlink[0]{}%
\providecommand \url  [0]{\begingroup\@sanitize@url \@url }%
\providecommand \@url [1]{\endgroup\@href {#1}{\urlprefix }}%
\providecommand \urlprefix  [0]{URL }%
\providecommand \Eprint [0]{\href }%
\providecommand \doibase [0]{http://dx.doi.org/}%
\providecommand \selectlanguage [0]{\@gobble}%
\providecommand \bibinfo  [0]{\@secondoftwo}%
\providecommand \bibfield  [0]{\@secondoftwo}%
\providecommand \translation [1]{[#1]}%
\providecommand \BibitemOpen [0]{}%
\providecommand \bibitemStop [0]{}%
\providecommand \bibitemNoStop [0]{.\EOS\space}%
\providecommand \EOS [0]{\spacefactor3000\relax}%
\providecommand \BibitemShut  [1]{\csname bibitem#1\endcsname}%
\let\auto@bib@innerbib\@empty
\bibitem [{\citenamefont {Seifert}(2008)}]{seifertprinciples}%
  \BibitemOpen
  \bibfield  {author} {\bibinfo {author} {\bibfnamefont {U.}~\bibnamefont
  {Seifert}},\ }\href@noop {} {\bibfield  {journal} {\bibinfo  {journal} {Eur.
  Phys. J. B}\ }\textbf {\bibinfo {volume} {64}},\ \bibinfo {pages} {423}
  (\bibinfo {year} {2008})}\BibitemShut {NoStop}%
\bibitem [{\citenamefont {Leff}\ and\ \citenamefont {Rex}(2003)}]{Leff03}%
  \BibitemOpen
  \bibfield  {author} {\bibinfo {author} {\bibfnamefont {H.~S.}\ \bibnamefont
  {Leff}}\ and\ \bibinfo {author} {\bibfnamefont {A.~F.}\ \bibnamefont {Rex}},\
  }\href@noop {} {\emph {\bibinfo {title} {Maxwell's Demon 2: Entropy,
  Classical and Quantum Information, Computing}}}\ (\bibinfo  {publisher}
  {Institute of Physics Publishing},\ \bibinfo {year} {2003})\BibitemShut
  {NoStop}%
\bibitem [{\citenamefont {Szilard}(1929)}]{Szilard29}%
  \BibitemOpen
  \bibfield  {author} {\bibinfo {author} {\bibfnamefont {L.}~\bibnamefont
  {Szilard}},\ }\href@noop {} {\bibfield  {journal} {\bibinfo  {journal} {Z. f.
  Physik}\ }\textbf {\bibinfo {volume} {53}},\ \bibinfo {pages} {840} (\bibinfo
  {year} {1929})}\BibitemShut {NoStop}%
\bibitem [{\citenamefont {Maxwell}(1871)}]{Maxwell71}%
  \BibitemOpen
  \bibfield  {author} {\bibinfo {author} {\bibfnamefont {J.~C.}\ \bibnamefont
  {Maxwell}},\ }\href@noop {} {\emph {\bibinfo {title} {Theory of Heat}}}\
  (\bibinfo  {publisher} {Longmans, Green and Co.},\ \bibinfo {year}
  {1871})\BibitemShut {NoStop}%
\bibitem [{\citenamefont {Earman}\ and\ \citenamefont
  {Norton}(1998)}]{Earman98}%
  \BibitemOpen
  \bibfield  {author} {\bibinfo {author} {\bibfnamefont {J.}~\bibnamefont
  {Earman}}\ and\ \bibinfo {author} {\bibfnamefont {J.~D.}\ \bibnamefont
  {Norton}},\ }\href@noop {} {\bibfield  {journal} {\bibinfo  {journal} {Stud.
  Hist. Phil. Mod. Phys.}\ }\textbf {\bibinfo {volume} {29}},\ \bibinfo {pages}
  {435} (\bibinfo {year} {1998})}\BibitemShut {NoStop}%
\bibitem [{\citenamefont {Earman}\ and\ \citenamefont
  {Norton}(1999)}]{Earman99}%
  \BibitemOpen
  \bibfield  {author} {\bibinfo {author} {\bibfnamefont {J.}~\bibnamefont
  {Earman}}\ and\ \bibinfo {author} {\bibfnamefont {J.~D.}\ \bibnamefont
  {Norton}},\ }\href@noop {} {\bibfield  {journal} {\bibinfo  {journal} {Stud.
  Hist. Phil. Mod. Phys.}\ }\textbf {\bibinfo {volume} {30}},\ \bibinfo {pages}
  {1} (\bibinfo {year} {1999})}\BibitemShut {NoStop}%
\bibitem [{\citenamefont {Hemmo}\ and\ \citenamefont
  {Shenker}(2012)}]{HemShen12}%
  \BibitemOpen
  \bibfield  {author} {\bibinfo {author} {\bibfnamefont {M.}~\bibnamefont
  {Hemmo}}\ and\ \bibinfo {author} {\bibfnamefont {O.}~\bibnamefont
  {Shenker}},\ }\href@noop {} {\emph {\bibinfo {title} {{The Road to Maxwell's
  Demon}}}}\ (\bibinfo  {publisher} {Cambridge},\ \bibinfo {year}
  {2012})\BibitemShut {NoStop}%
\bibitem [{\citenamefont {Brillouin}(1951)}]{Brillouin51}%
  \BibitemOpen
  \bibfield  {author} {\bibinfo {author} {\bibfnamefont {L.}~\bibnamefont
  {Brillouin}},\ }\href@noop {} {\bibfield  {journal} {\bibinfo  {journal} {J.
  Appl. Phys.}\ }\textbf {\bibinfo {volume} {22}},\ \bibinfo {pages} {334}
  (\bibinfo {year} {1951})}\BibitemShut {NoStop}%
\bibitem [{\citenamefont {Sagawa}\ and\ \citenamefont {Ueda}(2009)}]{Sagawa09}%
  \BibitemOpen
  \bibfield  {author} {\bibinfo {author} {\bibfnamefont {T.}~\bibnamefont
  {Sagawa}}\ and\ \bibinfo {author} {\bibfnamefont {M.}~\bibnamefont {Ueda}},\
  }\href@noop {} {\bibfield  {journal} {\bibinfo  {journal} {Phys. Rev. Lett.}\
  }\textbf {\bibinfo {volume} {102}},\ \bibinfo {pages} {250602} (\bibinfo
  {year} {2009})}\BibitemShut {NoStop}%
\bibitem [{\citenamefont {Granger}\ and\ \citenamefont
  {Kantz}(2011)}]{Granger11}%
  \BibitemOpen
  \bibfield  {author} {\bibinfo {author} {\bibfnamefont {L.}~\bibnamefont
  {Granger}}\ and\ \bibinfo {author} {\bibfnamefont {H.}~\bibnamefont
  {Kantz}},\ }\href@noop {} {\bibfield  {journal} {\bibinfo  {journal} {Phys.
  Rev. E}\ }\textbf {\bibinfo {volume} {84}},\ \bibinfo {pages} {061110}
  (\bibinfo {year} {2011})}\BibitemShut {NoStop}%
\bibitem [{\citenamefont {Mandal}\ and\ \citenamefont
  {Jarzynski}(2012)}]{Mandal12}%
  \BibitemOpen
  \bibfield  {author} {\bibinfo {author} {\bibfnamefont {D.}~\bibnamefont
  {Mandal}}\ and\ \bibinfo {author} {\bibfnamefont {C.}~\bibnamefont
  {Jarzynski}},\ }\href@noop {} {\bibfield  {journal} {\bibinfo  {journal}
  {Proc. Natl. Acad. Sci. USA}\ }\textbf {\bibinfo {volume} {109}},\ \bibinfo
  {pages} {11641} (\bibinfo {year} {2012})}\BibitemShut {NoStop}%
\bibitem [{\citenamefont {Sagawa}\ and\ \citenamefont
  {Ueda}(2012{\natexlab{a}})}]{Sagawa12}%
  \BibitemOpen
  \bibfield  {author} {\bibinfo {author} {\bibfnamefont {T.}~\bibnamefont
  {Sagawa}}\ and\ \bibinfo {author} {\bibfnamefont {M.}~\bibnamefont {Ueda}},\
  }\href@noop {} {\bibfield  {journal} {\bibinfo  {journal} {Phys. Rev. E}\
  }\textbf {\bibinfo {volume} {85}},\ \bibinfo {pages} {021104} (\bibinfo
  {year} {2012}{\natexlab{a}})}\BibitemShut {NoStop}%
\bibitem [{\citenamefont {Sagawa}\ and\ \citenamefont
  {Ueda}(2012{\natexlab{b}})}]{Sagawa12b}%
  \BibitemOpen
  \bibfield  {author} {\bibinfo {author} {\bibfnamefont {T.}~\bibnamefont
  {Sagawa}}\ and\ \bibinfo {author} {\bibfnamefont {M.}~\bibnamefont {Ueda}},\
  }\href@noop {} {\bibfield  {journal} {\bibinfo  {journal} {Phys. Rev. Lett.}\
  }\textbf {\bibinfo {volume} {109}},\ \bibinfo {pages} {180602} (\bibinfo
  {year} {2012}{\natexlab{b}})}\BibitemShut {NoStop}%
\bibitem [{\citenamefont {Ford}(2016)}]{Ford16}%
  \BibitemOpen
  \bibfield  {author} {\bibinfo {author} {\bibfnamefont {I.~J.}\ \bibnamefont
  {Ford}},\ }\href {\doibase 10.1080/ 00107514.2015.1121604} {\bibfield
  {journal} {\bibinfo  {journal} {Contemp. Phys.}\ } (\bibinfo {year} {2016}),\
  10.1080/ 00107514.2015.1121604}\BibitemShut {NoStop}%
\bibitem [{\citenamefont {Landauer}(1961)}]{Landauer61}%
  \BibitemOpen
  \bibfield  {author} {\bibinfo {author} {\bibfnamefont {R.}~\bibnamefont
  {Landauer}},\ }\href@noop {} {\bibfield  {journal} {\bibinfo  {journal} {IBM
  J. Res. Dev.}\ }\textbf {\bibinfo {volume} {5}},\ \bibinfo {pages} {183}
  (\bibinfo {year} {1961})}\BibitemShut {NoStop}%
\bibitem [{\citenamefont {Bennett}(1973)}]{Bennett73}%
  \BibitemOpen
  \bibfield  {author} {\bibinfo {author} {\bibfnamefont {C.~H.}\ \bibnamefont
  {Bennett}},\ }\href@noop {} {\bibfield  {journal} {\bibinfo  {journal} {IBM
  J. Res. Dev.}\ }\textbf {\bibinfo {volume} {17}},\ \bibinfo {pages} {525}
  (\bibinfo {year} {1973})}\BibitemShut {NoStop}%
\bibitem [{\citenamefont {Bennett}(1982)}]{Bennett82}%
  \BibitemOpen
  \bibfield  {author} {\bibinfo {author} {\bibfnamefont {C.~H.}\ \bibnamefont
  {Bennett}},\ }\href@noop {} {\bibfield  {journal} {\bibinfo  {journal} {Int.
  J. Theor. Phys.}\ }\textbf {\bibinfo {volume} {21}},\ \bibinfo {pages} {905}
  (\bibinfo {year} {1982})}\BibitemShut {NoStop}%
\bibitem [{\citenamefont {Plenio}\ and\ \citenamefont
  {Vitelli}(2001)}]{Plenio01}%
  \BibitemOpen
  \bibfield  {author} {\bibinfo {author} {\bibfnamefont {M.~B.}\ \bibnamefont
  {Plenio}}\ and\ \bibinfo {author} {\bibfnamefont {V.}~\bibnamefont
  {Vitelli}},\ }\href@noop {} {\bibfield  {journal} {\bibinfo  {journal}
  {Contemp. Phys.}\ }\textbf {\bibinfo {volume} {42}},\ \bibinfo {pages} {25}
  (\bibinfo {year} {2001})}\BibitemShut {NoStop}%
\bibitem [{\citenamefont {Abreu}\ and\ \citenamefont
  {Seifert}(2011)}]{Abreu11}%
  \BibitemOpen
  \bibfield  {author} {\bibinfo {author} {\bibfnamefont {D.}~\bibnamefont
  {Abreu}}\ and\ \bibinfo {author} {\bibfnamefont {U.}~\bibnamefont
  {Seifert}},\ }\href@noop {} {\bibfield  {journal} {\bibinfo  {journal} {Eur.
  Phys. Lett.}\ }\textbf {\bibinfo {volume} {94}},\ \bibinfo {pages} {10001}
  (\bibinfo {year} {2011})}\BibitemShut {NoStop}%
\bibitem [{\citenamefont {Barato}\ and\ \citenamefont
  {Seifert}(2013)}]{Barato13}%
  \BibitemOpen
  \bibfield  {author} {\bibinfo {author} {\bibfnamefont {A.~C.}\ \bibnamefont
  {Barato}}\ and\ \bibinfo {author} {\bibfnamefont {U.}~\bibnamefont
  {Seifert}},\ }\href@noop {} {\bibfield  {journal} {\bibinfo  {journal} {Eur.
  Phys. Lett.}\ }\textbf {\bibinfo {volume} {101}},\ \bibinfo {pages} {60001}
  (\bibinfo {year} {2013})}\BibitemShut {NoStop}%
\bibitem [{\citenamefont {Spinney}\ and\ \citenamefont
  {Ford}(2012)}]{SpinneyFord12b}%
  \BibitemOpen
  \bibfield  {author} {\bibinfo {author} {\bibfnamefont {R.~E.}\ \bibnamefont
  {Spinney}}\ and\ \bibinfo {author} {\bibfnamefont {I.~J.}\ \bibnamefont
  {Ford}},\ }\href@noop {} {\bibfield  {journal} {\bibinfo  {journal} {Phys.
  Rev. E}\ }\textbf {\bibinfo {volume} {85}},\ \bibinfo {pages} {051113}
  (\bibinfo {year} {2012})}\BibitemShut {NoStop}%
\bibitem [{\citenamefont {Ford}(2013)}]{fordbook}%
  \BibitemOpen
  \bibfield  {author} {\bibinfo {author} {\bibfnamefont {I.~J.}\ \bibnamefont
  {Ford}},\ }\href@noop {} {\emph {\bibinfo {title} {Statistical Physics: an
  entropic approach}}}\ (\bibinfo  {publisher} {Wiley},\ \bibinfo {year}
  {2013})\BibitemShut {NoStop}%
\end{thebibliography}

%

\end{document}